\documentclass[prb,superscriptaddress,longbibliography,twocolumn]{revtex4-2}

\usepackage{graphicx}
\usepackage{bm}
\usepackage{xfrac}
\usepackage{xcolor}
\usepackage{physics}
\usepackage{array}
\usepackage{amsmath}
\usepackage{amsfonts}
\usepackage{appendix}
\usepackage{float}
\usepackage{multirow}
\usepackage{tabularx}
\usepackage{comment}

\newcolumntype{P}[1]{>{\centering\arraybackslash}p{#1}}

\newcommand{\be}{\begin{equation}}
\newcommand{\ee}{\end{equation}}

\newcommand{\br}{{{\bf{r}}}}

\newcommand{\bea}{\begin{eqnarray}}
\newcommand{\eea}{\end{eqnarray}}
\newcommand{\beal}{\begin{align}}
\newcommand{\eeal}{\end{align}}

\newcommand{\upa}{\uparrow}
\newcommand{\dna}{\downarrow}

\newcommand{\vS}{{\bf S}}
\newcommand{\dg}{{\dagger}}
\newcommand{\pdg}{{\phantom\dagger}}

\newcommand{\tS}{{\tilde S}}

\newcommand{\bigJ}{\mathbb{J}}

\begin{document}

\title{Impact of gapped spin-orbit
excitons on low energy pseudospin exchange interactions}
\author{Sreekar Voleti}
\thanks{These authors contributed equally to this work}
\affiliation{Department of Physics, University of Toronto, 60 St. George Street, Toronto, ON, M5S 1A7 Canada}
\author{F. David Wandler}
\thanks{These authors contributed equally to this work}
\affiliation{Department of Physics, University of Toronto, 60 St. George Street, Toronto, ON, M5S 1A7 Canada}
\author{Arun Paramekanti}
\affiliation{Department of Physics, University of Toronto, 60 St. George Street, Toronto, ON, M5S 1A7 Canada}
\date{\today}

\begin{abstract}
The quest for exotic quantum magnetic ground states, including the Kitaev spin liquid and quantum spin-ices, 
has led to the discovery of several
quantum materials where low energy pseudospin-$1/2$ doublets 
arise from the splitting of spin-orbit entangled
multiplets with higher degeneracy. 
Such systems include
$d$-orbital and
$f$-orbital Mott insulators. 
When the gap between the low energy pseudospin-$1/2$ levels and 
the excited levels of the multiplet or `excitons' is not large,
the effective low-energy 
exchange interactions between the low energy pseudospin-$1/2$ 
moments can acquire significant corrections from coupling
to the excitons. We extract these corrections using higher order perturbation
theory as well as an exact Schrieffer-Wolff transformation. Such corrections can impact the exchange matrix for the low energy
pseudospin-$1/2$ levels by renormalizing
the strength and the sign of Heisenberg exchange or Ising anisotropies, and potentially even inducing
bond-anisotropic couplings such as Kitaev-$\Gamma$ exchange interactions. We discuss recent experiments on various
cobaltate and osmate materials which hint at the ubiquity and importance of this physics.
\end{abstract}
\maketitle


Magnetic solids exhibit strong quantum spin fluctuations 
in the limit of small spin.
Spin-$1/2$ systems are thus natural candidates to look for exotic phases of quantum matter 
including quantum spin liquids. The simplest realization of such spin-$1/2$ degrees
of freedom corresponds to single electrons nailed down at atomic sites in a single-orbital
Mott insulating crystal. The low energy ordering and dynamics of such single-orbital 
Mott insulators can be described using effective Heisenberg models in the limit of
strong Hubbard repulsion, with higher order
ring-exchange terms becoming important for moderate Hubbard repulsion. A prototypical example
is La$_2$CuO$_4$\cite{LCO_Anderson1987,LCO_Shirane1987,LCO_Chakravarty1989,LCO_Hayden1991,LCO_Coldea2001}, 
the undoped parent of the cuprate superconductors.

More interesting
realizations of low-spin quantum magnets occur in multi-orbital systems with spin-orbit
coupling (SOC), where the role of ``spin'' is played by an effective pseudospin-$1/2$ moment 
with entangled spin and orbital degrees of freedom. The most well-studied example of
this kind are the $j_{\rm eff}\!=\! 1/2$ Mott insulators in  compounds such as the layered Ir$^{4+}$ perovskite iridate
Sr$_{2}$IrO$_{4}$ \cite{SrIO_BJKim2008,SrIO_Jackeli2009,SrIO_Senthil2011,SrIO_Kim2012,SrIO_Min2012,SrIO_Fujiyama2012},
the layered honeycomb and hyperhoneycomb 
polytypes of A$_2$IrO$_3$ (with A=Li, Na)\cite{iridates_kimchi2014,Iridates_review_Trebst2022,KitaevReview_Takagi2019}, or the analogue
Ru$^{3+}$ honeycomb ruthenate $\alpha$-RuCl$_3$\cite{RuCl3_SOC_Kim2014,RuCl3_neutron_Nagler2017}. In these
cases, SOC splits the six-fold degenerate local $t_{2g}$ orbitals
(including spin) into a lower $j_{\rm eff}\!=\! 1/2$ doublet with 
a significant gap $\sim\!0.2$-$0.6$\,eV to the higher energy 
$j_{\rm eff}\!=\! 3/2$ quartet which has been termed a 
`spin-orbit exciton'.

Here, we will focus on a distinct class of 
interesting pseudospin-$1/2$
magnets which appear in a variety of 
$d$-orbital transition metal oxides,
and $f$-orbital heavy fermion materials, 
where the pseudospin
doublet arises from weak splitting of a higher moment multiplet with SOC. 
A simple illustrative example is the case of a 
spin-$3/2$ multiplet which splits
into a pair of Kramers doublets with $S_z\!=\! \pm 1/2$ and
$S_z\!=\! \pm 3/2$ due to SOC in a tetragonal crystal. In this case,
the lower Kramers doublet acts as a low-energy pseudospin-$1/2$ degree 
of freedom while the upper doublet may be viewed as a `gapped exciton'.
However, the exciton gap is not large.
In order to understand the low energy emergent quantum phases of these 
pseudospin-$1/2$ magnets, we have
to first extract the effective Hamiltonian describing 
the interaction between these doublets. This is
commonly done by appealing to microscopic calculations 
of the two-site exchange 
interaction between the pseudospin-$1/2$ moments
(e.g.,
from tight-binding models based on density functional theory), 
or tuning parameters of 
symmetry-based model spin Hamiltonians
to fit experimental data from low energy probes such as inelastic 
neutron scattering. The reduction of the Hamiltonian from the
full Hilbert space to 
the low energy pseudospin-$1/2$ Hilbert space is
important to enable numerical studies on larger system sizes.

A key message of our work is that in Mott insulators
where the splitting $\Delta$ between the low energy 
pseudospin-$1/2$
doublet and the `gapped exciton' is not very large, the
correct way to extract the two-site pseudospin exchange 
starting from an electronic Hamiltonian
is via a two-step procedure. The
first step involves second-order perturbation theory in the
electron hopping which couples the entire pair of
nearest neighbor multiplets. In Mott-Hubbard insulators,
this results in a matrix of exchange
couplings with an exchange scale $J_{\rm ex}
\!\propto\! t^2/U$ where $t$ is used as a shorthand for 
the orbital-dependent electron hopping matrix elements, and $U$ is used as a shorthand for scales arising from Kanamori interactions.
The second step is to integrate
out the higher levels of the multiplet, which are split off
by $\Delta$, leading to an 
effective pseudospin-$1/2$ model. This induces important
exchange corrections which are on the scale of $J_{\rm ex}^2/\Delta$
which is thus fourth-order in the electron hopping.
We will discuss several examples showing how the
resulting low energy effective Hamiltonian can differ
significantly from the naive result where we project to the low
energy doublet from the outset. 

Quantum magnets which possess a pair of weakly split
Kramers doublets can be realized in several octahedrally 
coordinated Mott insulators with SOC, 
so it is not an uncommon scenario. Examples of such systems 
include $d^7$ cobaltates such as CoTiO$_3$ which exhibits 
low energy Dirac magnons and dispersive spin-orbit excitons \cite{CTO_Yuan2020,CTO_Coldea_Ncomm2021,VoletiD72021},
and candidate Kitaev materials such as BaCo$_2$(AsO$_4$)$_2$,  BaCo2(PO$_4$)$_2$, 
Na$_3$Co$_2$SbO$_6$ and Na$_2$Co$_2$TeO$_6$\cite{BCAO_Regnault2018, BCPO_Nair2018, ncso_mcguire2019,ncto_ncso_stock2020,ncto_simonet2016,BCAO_Broholm2023,Armitage2022,VoletiD72021} These systems
with strong trigonal distortion realize an effective
spin $S\!=\!3/2$ moment which is split into two Kramers 
doublets by SOC.
Other examples include $d^1$ Mott insulators such as Ba$_2$MgReO$_6$
which displays a higher temperature quadrupolar and lower
temperature dipolar magnetic ordering transitions,
and magnetically ordered $d^3$ materials such as Sr$_2$FeOsO$_6$ \cite{Hiroi_JPSJ2019,Hiroi_PRR2020}
In these systems, the pair of Kramers doublets may arise,
respectively,
from tetragonal splitting of a $j\!=\!3/2$ or $J\!=\!3/2$ moment.


A distinct type of weakly split multiplet is
realized $d^2$ Mott insulators which host an angular momentum $J\!=\! 2$ multiplet 
that splits into a ground non-Kramers $E_g$ pseudospin-$1/2$
doublet and an excited $T_{2g}$ triplet even in an octahedral crystal field. 
Recent work has revealed osmate double perovskites such
as Ba$_2$MOsO$_6$ (M = Zn, Mg, Ca) as candidates for realizing 
such non-Kramers doublets\cite{Maharaj2019,MaharajOct2020,ParamekantiOct2020,VoletiOct2020,VoletiOct2021}
In this case, the low energy $\tau_{x}$ and $\tau_{z}$ pseudospin operators transform 
as a two-component electric quadrupole, while $\tau_y$ transforms 
as an Ising magnetic octupole. These compounds appear to show some evidence
for ferro-octupolar ordering of the non-Kramers doublets, while the higher energy
$T_{2g}$ triplet acts as a `gapped exciton'. In this case, the small
$E_g$-$T_{2g}$ exciton gap arises due to a combination of Hund's coupling and 
SOC-induced virtual transitions from single-particle $t_{2g}$ to $e_g$ levels.

We will discuss several models where
the coupling between the lower 
and upper multiplet significantly impacts the naive
low-energy Hamiltonian. Using a two-step perturbation theory,
we show that this can renormalize and even
potentially flip the sign of the exchange
couplings, or can generate entirely new bond-anisotropic
terms such as Kitaev or off-diagonal
$\Gamma$ interactions. 
We test our two-step perturbative results against
an exact Schrieffer-Wolff transformation.
We note that similar ideas have
also been explored in recent work with
applications to Sr$_2$IrO$_4$, and may also be
relevant to anisotropic and higher-order
spin interactions in heavy fermion systems.

\section{Extended Perturbation Theory}
Let us consider a $D$-dimensional multiplet at each site 
split by energy $\Delta$ 
into low energy `pseudospin' multiplet of degeneracy $D_L$ and a 
high energy `exciton' multiplet of degeneracy $D_H\!=\!D\!-\!D_L$.
For the case of spin-$3/2$ split into two Kramers doublets,
$D\!=\!4$ and $D_L\!=\!D_H\!=\!2$. For the $d^2$ ion split
into a non-Kramers pseudospin and a triplet exciton, we have 
$D\!=\!5$ with $D_L\!=\!2$ and $D_H\!=\!3$.
When a neighboring pair of sites are connected by a hopping Hamiltonian $H_T$, the standard procedure for computing the 
two-site pseudospin exchange
involves treating $H_T$ within second order perturbation theory, 
integrating out the intermediate charge transfer excitations 
which are at much 
higher energy $\sim\!U$ (the Hubbard interaction).
This leads to a $D_L^2 \times D_L^2$ Hamiltonian matrix which can be recast
in terms of exchange interaction parameters between the 
pseudospins.
However, when $\Delta$ is small, in a manner to be clarified 
below, the correct procedure is a two-step approach. The
first step is to extract the full
$D^2 \times D^2$ Hamiltonian $\mathcal{V}_J$ which
espouses all second order contributions in $H_T$
to exchange couplings between the entire $J$-multiplets 
(i.e., both pseudospins and 
excitons). The second step is to integrate out the high energy
excitons and obtain an effective low-energy pseudospin Hamiltonian.
Accordingly, we split up the full two-site multiplet Hamiltonian, obtained 
at the end of the first step above, as 
$   \mathcal{H} = H_0 + \mathcal{V}_J$,
where $H_0$ represents the on-site splitting $\Delta$
between the pseudospin and exciton levels, and $V_J$ is
${\cal O}(t^2/U)$.
This site-localized Hamiltonian $H_0$ has three distinct 
energy 
levels:
(i) $E^{(0)}_0$ corresponding to both sites being in 
the pseudospin branch, (ii)
$E^{(0)}_1\!=\!E^{(0)}_0\!+\!\Delta$ 
corresponding to one of the sites being in
the exciton branch, and (iii)
$E^{(0)}_2\!=\! E^{(0)}_0\!+\! 2\Delta$ 
when both sites live in the exciton branch. 
The degeneracies of these levels are $D_L^2$, $2 D^\pdg_L D^\pdg_H$, and $D_H^2$ respectively. 
Typically, the effective Hamiltonian between the sites 
is just extracted at ${\cal O}(t^2/U)$ as the projection of $\mathcal{V}_J$ onto the $E^{(0)}$ manifold i.e. 
    $H_{\rm eff}^{[1]} \!=\! P_0 \, \mathcal{V}_J \, P_0$,
where $P_0$ is the projector onto the $E^{(0)}$ subspace. 
The exciton-induced correction is given by 
\begin{equation} \label{ptexchange}
H_{\rm eff}^{[2]} = P_0 \, \mathcal{V}_J \,P_1 \left( \frac{1}{E^{(0)}_0-H_0} \right) P_1 \, \mathcal{V}_J \, P_0
\end{equation}
where $P_1 \!=\! 1\!-\! P_0$; this expression in Eq.~\eqref{ptexchange} is {\it fourth} order in the hopping Hamiltonian $H_T$ 
between the sites, and is typically ignored. While this 
term is $\sim\!{\cal O}(t^4/U^2\Delta)$, it can nevertheless become 
comparable to the conventional exchange coupling, when $\Delta \!\sim\! {\cal O}(t^2/U)$. 

\subsection{ Split $J=3/2$ moment }
Here, we apply the extended perturbation theory to an effective split $J = 3/2$ system (i.e. with $D_L = D_H = 2$). Before exploring the physics, we establish a useful basis for the two-site problem.
Using $\sigma^a$ to denote the usual Pauli matrices \footnote{We take the convention that $\sigma^0$ is the $2\times2$ identity matrix and $\sigma^1 = \sigma^x$, $\sigma^2 = \sigma^y$, and $\sigma^3 = \sigma^z$.}, we define the following convenient basis for the $4\times4$ Hermitian matrices (written in the basis $\{\ket{1/2}, \ket{-1/2}, \ket{3/2}, \ket{-3/2}\}$) that can act on each site:
\begin{equation}
\label{eqn:basisdef1}
\eta^a = \begin{pmatrix} \sigma^a & 0 \\ 0 & 0 \end{pmatrix} \quad , \quad \tau^a =  \begin{pmatrix} 0 & 0 \\ 0 & \sigma^a  \end{pmatrix},
\end{equation}
\begin{equation}
\label{eqn:basisdef2}
\xi_r^a = \frac{1}{\sqrt{2}} \begin{pmatrix} 0 & \sigma^a \\  \sigma^a & 0  \end{pmatrix}\quad , \quad \xi_i^a = \frac{1}{\sqrt{2}} \begin{pmatrix} 0 & -i\sigma^a \\  i\sigma^a & 0  \end{pmatrix}.
\end{equation} 
Here, $a\in \{0,1,2,3\}$ for each type of operator. In this basis, the $\eta^a$ operate within the $J_z^2=1/2$ subspace, the $\tau^a$ operate within the $J_z^2=3/2$ subspace, and the $\xi_r^a$ and $\xi_i^a$ swap states between these two subspaces. 
These 16 matrices are more convenient than the usual $J=3/2$ multipole operator basis, because they naturally separate the two doublets. See Appendix \ref{app:bases} for the change of basis to multipole operators.

For the purposes of integrating out the excitons to leading order, we only need the following terms from the $J=3/2$ interaction Hamiltonian:
\begin{equation}
    \begin{split}
        \mathcal{V}_J = J^{(\eta\eta)}_{ab} & \eta^a \eta^b + \sum_{s = r,i} K^{(s)}_{ab}\left(\eta^a\xi^b_s + \xi^b_s\eta^a\right) \\ 
        + & \sum_{s,t = r,i} M^{(st)}_{ab} \xi^a_s\xi^b_t
    \end{split}
\end{equation}
These three terms correspond to processes that do not excite an exciton, excite exactly one exciton, and excite two excitons, respectively. Other terms would annihilate states with no excitons, so they cannot contribute to the physics of the lower doublet at second order in perturbation theory (i.e. they would vanish when taking the projection via $P_0$ to the lower energy sector in Equation \ref{ptexchange}).

We are looking for an interaction matrix between the pseudospin-1/2 moments, which we call $J_{\rm eff}$. This should be understood as the Hamiltonian

\begin{equation}
    H_{\rm eff} = \tilde{\mathbf{s}}_1^T \text{ } J_{\rm eff} \text{ } \tilde{\mathbf{s}}_2
\end{equation}

where $\tilde{\mathbf{s}}$ refers to the vector of spin operators in the two dimensional pseudospin space, and the numbered subscript indicates the site index. The second order perturbation calculation can be done according to Equation \eqref{ptexchange}, yielding:
\begin{equation}
J_{\rm eff} = J^{(\eta \eta)} + \delta J_K + \delta J_M
\end{equation}
where 
\begin{equation}
\label{KHamPT}
\begin{split}
(\delta J_K)_{ab} = -\frac{1}{2\Delta} & \left(K^{(r)}_{cd} - i K^{(i)}_{cd}\right)\left(K^{(r)}_{ef} + i K^{(i)}_{ef}\right)\\ & \times \left(\lambda_{cea}\lambda_{dfb} + \lambda_{ceb}\lambda_{dfa}\right)
\end{split}
\end{equation}
and
\begin{equation}
\label{MHamPT}
\begin{split}
(\delta J_M)_{ab} = -\frac{1}{8\Delta}  & \left(M^{(rr)}_{cd} - M^{(ii)}_{cd} - i M^{(ri)}_{cd} - i M^{(ir)}_{cd} \right)\\ &\times \left(M^{(rr)}_{ef} - M^{(ii)}_{ef} + i M^{(ri)}_{ef} + i M^{(ir)}_{ef} \right) \\ &\times \lambda_{cea}\lambda_{dfb}
\end{split}
\end{equation}
Here we have defined the lambda symbol by $\sigma^a\sigma^b = \lambda^{abc}\sigma^c$. Explicitly,
    \begin{equation}
\lambda_{abc} = \begin{cases} \delta_{bc} & a=0 \\ \delta_{ac} & b=0 \\ \delta_{ab} & c=0 \\ i \varepsilon_{abc} & a,b,c\neq 0\end{cases}.
\end{equation}



It is clear from these equations that having $K \sim \sqrt{J^{(\eta\eta)}\Delta}$ or $M \sim \sqrt{J^{(\eta\eta)}\Delta}$ could lead to changes on the order of $J^{(\eta\eta)}$. In Section IV, we will demonstrate some toy examples where this occurs and completely changes the physics of the resulting spin theory. For now we test  extended perturbation theory on physically realistic models.

\subsection{First application: Spin-3/2 with tetragonal distortion}
Let us consider a ${\tilde S}\!=\!3/2$ multiplet, where this large
`spin' might experience weak SOC, or arise as a strongly spin-orbit coupled
$j\!=\! 3/2$ or $J\!=\! 3/2$ moment. We assume this is
split into two Kramers doublets via a 
tetragonal distortion encapsulated by the Hamiltonian 
\begin{equation}
    H_0 = \Delta \left[ \tau_0(\br) + \tau_0(\br') + 
    2 \tau_0(\br) \tau_0(\br') \right].
\end{equation}
Here, $\tau^0 = (Q_{z^2} + 1)/2$ with $Q_{z^2}=\tilde{S}_z^2-\tilde{S}(\tilde{S}+1)/3$.
Let $\mathcal{V}_J$ contain Heisenberg spin exchange, as well as quadrupole, and 
octupole interactions given by
\begin{eqnarray} 
\label{sm32_tetragonal}
    \mathcal{V}_J &=& \bigJ_H \, \tilde{\mathbf{S}}(\br) \cdot \tilde{\mathbf{S}}(\br') + \bigJ_Q \, Q_{xy}(\br) Q_{xy}(\br')
    \nonumber \\
    &+& \bigJ_T \, T_{xyz}(\br) T_{xyz}(\br') 
\end{eqnarray}
where $Q_{xy} = (\tS_x \tS_y + \tS_y \tS_x)/\sqrt{3}$ 
is the quadrupole operator,
and $T_{xyz} = 2 {\rm Sym}[\tS_x \tS_y \tS_z]/3\sqrt{3}$
is the Ising-like octupole operator with ``Sym'' denoting
symmetrization.
Let us denote pseudospin-$1/2$ operators acting on the
low energy doublet as
$\tilde{s}_\alpha=\sigma_\alpha/2$ where $\sigma$ are Pauli matrices.
A simple projection of the spin-$3/2$ Hamiltonian into this pseudospin-$1/2$ doublet 
leads to
\begin{equation}
H_{\rm eff , ex}^{[1]} = \bigJ_H \left( \tilde{s}_x^1 \tilde{s}_x^2 + \tilde{s}_y^1 \tilde{s}_y^2 \right) 
+  \dfrac{\bigJ_{H}}{4} \tilde{s}_z^1 \tilde{s}_z^2.
\end{equation}
which is completely devoid of any terms which include the impact 
of higher multipole interactions.
However, using the extended perturbation theory result in 
Eq.~\eqref{ptexchange}, we find
\begin{equation}
\begin{split}
H_{\rm eff , ex}^{[2]} = \left( \bigJ_H - \dfrac{ 3\bigJ_H(\bigJ_Q-\bigJ_T) }{ 4 \Delta } \right) \left( \tilde{s}_x^1 \tilde{s}_x^2 + \tilde{s}_y^1 \tilde{s}_y^2 \right) \\
+ \left( \dfrac{\bigJ_H}{4} - \dfrac{39 \bigJ_H^2}{16 \Delta} - \dfrac{\bigJ_Q \bigJ_T}{\Delta} \right)\tilde{s}_z^1 \tilde{s}_z^2
\end{split}
\end{equation}
From the above, we can see that while the form of the couplings is the same, the coupling strengths have the potential 
of being strongly renormalized by the presence of the exciton if the multipole couplings $\bigJ_Q, \bigJ_T \!\sim\! \Delta$ or
$\bigJ_Q \bigJ_T/\bigJ_H \!\sim\! \Delta$. The right 
combination of the multipole couplings
can strongly suppress the $zz$ interaction, giving rise to a 
pure $XY$ model, or even flip the sign of the XXZ anisotropy.

\section{Application to Microscopic calculations}
Here we provide examples of how the above protocol may be used in a typical microscopic calculation, and see how it produces markedly different results compared to the standard treatment outlined at the beginning of the previous section. We consider two cases: a $d^1$ honeycomb system subject to trigonal distortion, and a $d^2$ fcc system that hosts higher order multipole moments in its ground state. These cases both feature larger moments that are split to give a (pseudo)spin-1/2 ground state, are numerically tractable via an exact Schrieffer-Wolff transformation (as outlined in Ref. \cite{VoletiOct2021}) in order to assess the accuracy of the EPT approach. The microscopic Hamiltonian for both of the cases is 
\bea
H_{\rm loc}=H_{\rm CEF}+H_{\rm SOC}+H_{\rm int}
\label{eq:hsingle}
\eea
which includes $t_{2g}$-$e_g$
crystal field splitting, SOC, and electronic interactions, written in the orbital basis 
($\{yz,xz,xy\},\{x^{2}\!-\!y^{2},3z^{2}\!-\!r^{2}\}$).
The CEF term is given by:
\begin{equation} \label{cfham}
H_{\rm CEF}=\sum_{\alpha, \beta} \sum_{s} A_{\alpha\beta}c^\dagger_{\alpha s} c_{\beta s}
\end{equation}
where $A$ is the local crystal field matrix written in the orbital basis, and $s$ is the spin. The SOC term is of the one-body form:
\begin{align}\label{eq:Hint}
\begin{split}
H_{\rm SOC} &= \frac{\lambda}{2}\sum_{\alpha, \beta} \sum_{s,s'} \bra{\alpha}\mathbf{L}\ket{\beta} \cdot \bra{s}\pmb{\sigma}\ket{s'}c^\dagger_{\alpha s} c_{\beta s'} \ ,
\end{split}
\end{align}
where $\pmb{\sigma}$ refers to the vector of Pauli matrices, and $\mathbf{L}$ are orbital angular momentum matrices. The operators $c_{\alpha s}$, $c^\dagger_{\alpha s}$ and $n_{\alpha s}$ destroy, create, and count the electrons with spin $s$ in orbital $\alpha$. The Kanamori interaction is given by
\bea
H_{\rm int} &=& U\sum_{\alpha}n_{\alpha \uparrow}n_{\alpha \downarrow} \!+\! \left( U' - {J_H \over 2} \right)  \sum_{\alpha > \beta} n_\alpha n_\beta 
 \\
 \!&-&\! J_H \sum_{\alpha \neq \beta} \vS_\alpha \cdot \vS_\beta \nonumber 
+ J_H \sum_{\alpha\neq\beta} c^\dg_{\alpha \upa} c^\dg_{\alpha\dna} c^\pdg_{\beta \downarrow} c^\pdg_{\beta \upa}
\eea
where $U$ and $U'$ are the intra- and inter-orbital Hubbard interactions, $J_H$ is the Hund's coupling, and $\vS_\alpha = (1/2) c^\dg_{\alpha s} \pmb{\sigma}_{s,s'} c^\pdg_{\alpha s'}$. The operator $n_\alpha\equiv n_{\alpha\uparrow}+n_{\alpha\downarrow}$ counts the total number of electrons in orbital $\alpha$. The spherical symmetry of the Coulomb interaction sets $U' = U - 2 J_H$ \cite{Georges2013}.

\subsection{$d^1$ ions in a honeycomb lattice}
\begin{figure*}
    \centering
    \includegraphics[width=\textwidth]{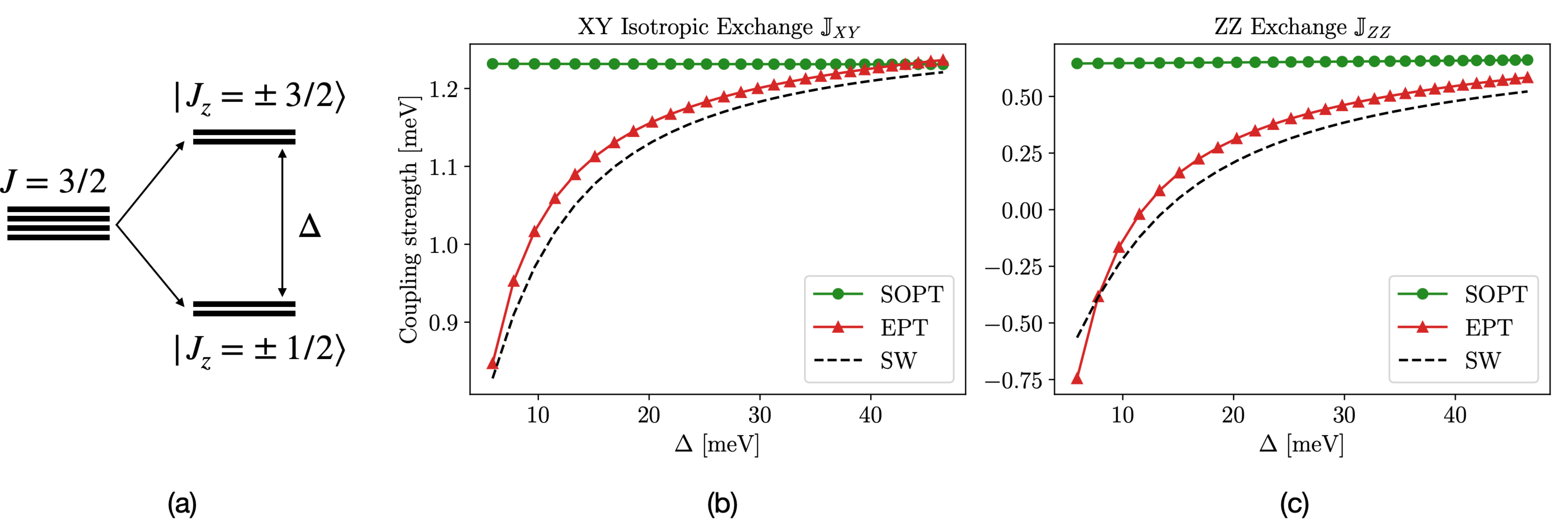}
    \caption{(a) Level structure for a $d^1$ ion with spin-orbit coupling and trigonal distortion. The lower $\ket{\pm 1/2}$ states act as the effective spin-1/2 moment. (b),(c) Exchange couplings for the spin Hamiltonian in Eq. \eqref{eq:d1spinham}, computed using SOPT and EPT, and compared to the exact SW calculation.}
    \label{fig:d1Fig}
\end{figure*}
A single-ion ground state of the Hamiltonian in Eq. \ref{eq:hsingle} with a single electron restricted to the $t_{2g}$ sector, is a four-fold degenerate $J=3/2$ manifold. A typical situation that arises in 2D materials is when this ion is in an octahedral cage, and the octahedra are used to form a honeycomb lattice. A natural distortion axis for such a lattice is that along the octahedral [111] direction, corresponding to the direction perpendicular to the honeycomb plane. Such a distorted octahedron has, in addition to the usual $t_{2g}$-$e_g$ splitting, the following term in the crystal field matrix:
$$ \frac{\delta}{3} \; (L_x + L_y + L_z)^2$$
where $\delta$ is the distortion parameter. Restricting ourselves to the $t_{2g}$ sector, the $A$ matrix is given by
\begin{eqnarray}
\left( \begin{array}{c c c}
        0 & \delta & \delta  \\
         \delta & 0 & \delta  \\
        \delta & \delta & 0  \\
    \end{array} \right). 
\end{eqnarray}
The effect of this distortion term is the split the $J=3/2$ moment into two Kramers doublets, with $\ket{\pm 1/2}$ as the ground state doublet, and $\ket{\pm 3/2}$ the `exciton', higher in energy by $\Delta$, as shown in Fig. \ref{fig:d1Fig}(a). To obtain the pseudospin exchange, we consider a two site model of such octahedra, connected via a hopping Hamiltonian of the form 
\bea
\label{hopham}
H_T^\gamma = \sum_{\alpha \beta s}(T^{\gamma}_{\alpha \beta} c^\dg_{2\beta s} c^\pdg_{1\alpha s} + T^{\gamma \dagger}_{\beta \alpha} c^\dg_{1\alpha s} c^\pdg_{2\beta s })
\eea
where $T^{\gamma}$ is the hopping matrix for the $\gamma$ bond. We consider a matrix for the $z$ bond in the honeycomb inspired by the 90 degree bonding geometry in Ref. \cite{JackeliKhaliullin2009}:
\bea
T^{z} = 
\left( \begin{array}{ccc}
0 & t_1 & 0  \\
t_1 & 0 & 0  \\
0 & 0 & t_2  \\
\end{array} 
\right)\ .
\eea
Here, $t_1$, is the $yz$-$zx$ hopping, and $t_2$ is the $xy$-$xy$ hopping. The matrices for the $x$ and $y$ bonds can be obtained via $C_3$ rotation about the octahedral [111] axis. For the illustrative case, we consider $t_1 = -100$ meV, and $t_2=50$ meV, along with the single ion parameters ($\lambda,\;U,\;J_H$) = $(0.1,\;2.5,\;0.3)$ eV. 
In the lab frame (see SI for details), the low energy pseudospin exchange matrix takes an XXZ form:
\bea 
\label{eq:d1spinham}
H_{\rm spin} = \mathbb{J}_{XY} \left( \tilde{s}_x^1 \tilde{s}_x^2 + \tilde{s}_y^1 \tilde{s}_y^2 \right) 
+  \mathbb{J}_{ZZ} \tilde{s}_z^1 \tilde{s}_z^2.
\eea
Figure \ref{fig:d1Fig}(b-c) shows the values of these exchange parameters when calculated using EPT, contrasted with the conventional method of directly projecting down to the lower manifold (SOPT). The two approaches are also compared with the exact two site Schrieffer-Wolff calculation. It can be seen that the EPT is much closer to the exact calculation, and the methods give significantly different coupling values. While the SOPT Hamiltonian remains XXZ for all gap values, it can be seen that for $\Delta \sim 17$ meV, the spin Hamiltonian is actually a pure XY model. It can also be seen that for a small enough gap value, we approach a point where $\mathbb{J}_{XY} \approx - \mathbb{J}_{ZZ}$. At this point, performing a single sublattice spin rotation such that ($\tilde{s}_x \rightarrow \tilde{s}_x$, $\tilde{s}_y \rightarrow -\tilde{s}_y$, $\tilde{s}_z \rightarrow -\tilde{s}_z$) would convert this into a pure Heisenberg antiferromagnet. Thus, the addition of the exciton mixing terms reveals a much richer class of spin Hamiltonians accessible via tuning the trigonal distortion.

\subsection{$d^2$ ions in an fcc lattice }
\begin{figure*}
    \centering
    \includegraphics[width=\textwidth]{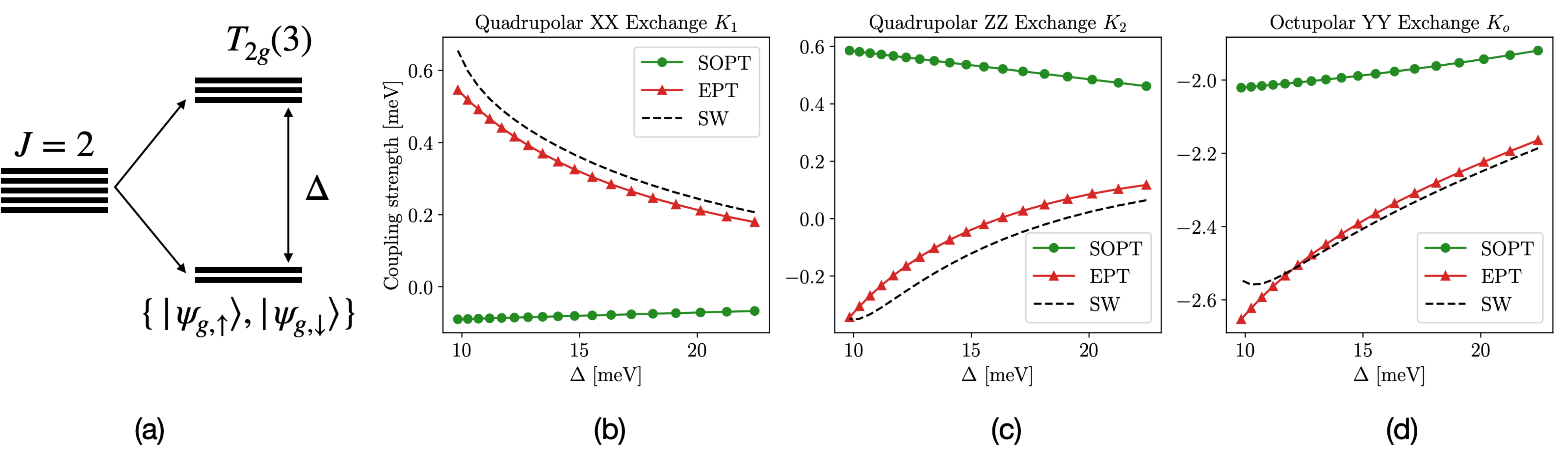}
    \caption{(a) Level structure for a $d^2$ ion within a double perovskite crystal. (b),(c),(d) Exchange couplings for the pseudospin Hamiltonian in Eq. \eqref{eq:d2spinham}, computed using SOPT and EPT, and compared to the exact SW calculation.}
    \label{fig:d2Fig}
\end{figure*}
Another class of systems where this formalism is useful is those where the pseudospin degree of freedom is made up of non-Krammers states. These have recently been studied in the context of $d^2$ Double Perovskites, where a $J\!=\! 2$ moment, when placed in a cubic environment, splits as $2 (E_g) \oplus 3 (T_{2g})$. The non-Kramers $E_g$ ground state may be treated as a pseudospin 1/2 degree of freedom, with wavefunctions 
\bea
|\psi_{g,\uparrow}\rangle =  \frac{1}{\sqrt{2}} (|2\rangle + | -2 \rangle);~~~
|\psi_{g,\downarrow}\rangle = |0\rangle.
\eea
Within this non-Kramers doublet space, the Pauli matrices $\tau_x,\tau_y,\tau_z$ are proportional to multipole operators, 
and are given by
$\tau_x \!\equiv\! (J_x^2\!-\!J_y^2)/2\sqrt{3}$,
$\tau_y \! \equiv\! \overline{J_x J_y J_z}/6\sqrt{3}$,
and $\tau_z \!\equiv\! (3 J_z^2\!-\! J(J+1))/6$, with overline denoting symmetrization.
Here, $\tau_x,\tau_z$
are electric quadrupoles while $\tau_y$ is a magnetic octupole. \cite{VoletiOct2020}. The form of the pseudospin Hamiltonian has been shown to take the form 
\begin{align}
\label{eq:d2spinham}
\!\! H_{\rm spin} \!=&\!\! \sum_{\langle i,j\rangle}\! \left[ K_{\rm o} \tau_{i y} \tau_{j y}
\!+\! \left( K_1 \cos^2\!\phi_{ij} \!+\! K_2 \sin^2\!\phi_{ij} \right) \tau_{i x} \tau_{j x} \right. \nonumber \\ 
 &+ \left( K_1 - K_2  \right) \sin\phi_{ij} \cos\phi_{ij} \left( \tau_{ix}\tau_{jz} + \tau_{iz}\tau_{jx} \right) \nonumber \\ 
 &+\left.  \left( K_1 \sin^2\phi_{ij} + K_2 \cos^2\phi_{ij}  \right)\tau_{i z} \tau_{j z}\right]
\end{align}
where $\phi_{ij} = \{ 0 , 2\pi/3 , 4\pi/3 \}$ correspond to nearest neighbors $(i,j)$ in the $\{ xy , yz , zx \}$ planes.
$K_{\rm o}$ and $K_{1,2}$ respectively correspond to the octupolar exchange and quadrupolar couplings.
An exact two-site calculation using a Schrieffer-Wolff transformation to obtain the effective low energy Hamiltonian indicated that the nearby $T_{2g}$ triplet is able to strongly influence the exchange parameters of the $E_g$ doublets. This system thus provides with another testing ground for the EPT formalism. As shown in Figure \ref{fig:d2Fig}(b-d), it can be seen that the dominant octupole-octupole exchange coupling shows a significant increase in magnitude, while also showing that the quadrupolar $K_1$ coupling is has the opposite sign and significantly higher magnitude compared to the SOPT case.  

\section{Some Interesting toy examples}
\label{sec:ToyModels}
In addition to the above physically motivated examples, it is important to note that this extended perturbation theory can lead to wildly different physics from the naive second order predictions. In the case of a two doublet system, any conceivable change in spin models, $\delta J$, can be realized with time-reversal invariant couplings between the doublets, with coupling coefficients in an intermediate scale between those of $\delta J$ and $\Delta$. An explicit proof of this is given in Appendix \ref{app:A2A} in the form of an algorithm that works backwards: taking any given $\delta J$ and working out a set of time-reversal invariant couplings that produce this $\delta J$ under perturbation theory. The system of equations that the algorithm solves is underdetermined meaning the results of this algorithm are not unique.

In the following subsections, we look at a few particularly striking cases with clean solutions. These demonstrate the power of inter-doublet couplings in changing the low energy physics.

\subsection{Changing the Heisenberg coupling}
To begin we consider the case where naive second order perturbation theory gives a Heisenberg (anti-)ferromagnet and the extended perturbation theory changes the strength or even the sign of the interaction. Hence, our starting spin model is 
\begin{equation}
    J = \begin{pmatrix}
    \mathcal{J}&0&0\\
    0&\mathcal{J}&0\\
    0&0&\mathcal{J}
    \end{pmatrix},
\end{equation}
with a correction of the form
\begin{equation}
\label{eqn:Heisenberg}
    \delta J = \begin{pmatrix}
    \kappa & 0 & 0 \\ 
    0 & \kappa & 0 \\
    0 & 0 & \kappa \\
    \end{pmatrix}.
\end{equation}
Such a correction can be introduced using only $K^{(i)}$ couplings. For $\kappa > 0$, this can be achieved by introducing 
\begin{equation}
    K^{(i)} = \begin{pmatrix}
    0 & 0 & 0 & 0 \\
    0 & \sqrt{\frac{\kappa \Delta }{2}} & 0 & 0 \\ 
    0 & 0 & \sqrt{\frac{\kappa \Delta }{2}} & 0 \\ 
    0 & 0 & 0 & \sqrt{\frac{\kappa \Delta }{2}} \\ 
    \end{pmatrix},
\end{equation}
and for $\kappa < 0$, this can be achieved by introducing 
\begin{equation}
    K^{(i)} = \begin{pmatrix}
    \sqrt{-2\kappa\Delta} & 0 & 0 & 0 \\
    0 & \sqrt{\frac{-\kappa \Delta }{2}} & 0 & 0 \\ 
    0 & 0 & \sqrt{\frac{-\kappa \Delta }{2}} & 0 \\ 
    0 & 0 & 0 & \sqrt{\frac{-\kappa \Delta }{2}} \\ 
    \end{pmatrix}.
\end{equation}
Notice here that to completely reverse the sign of the interaction (and therefore change the physics from a ferromagnet to an antiferromagnet or vice versa), we need $\kappa = -2\mathcal{J}$, so the $K^{(r)}$ couplings introduced are on the order of $\sqrt{\left|\mathcal{J}\right|\Delta}$ which is the geometric mean of the spin interaction scale $\mathcal{J}$ and the splitting scale $\Delta$.

\begin{figure*}
    \centering
    \includegraphics[width=0.7\textwidth]{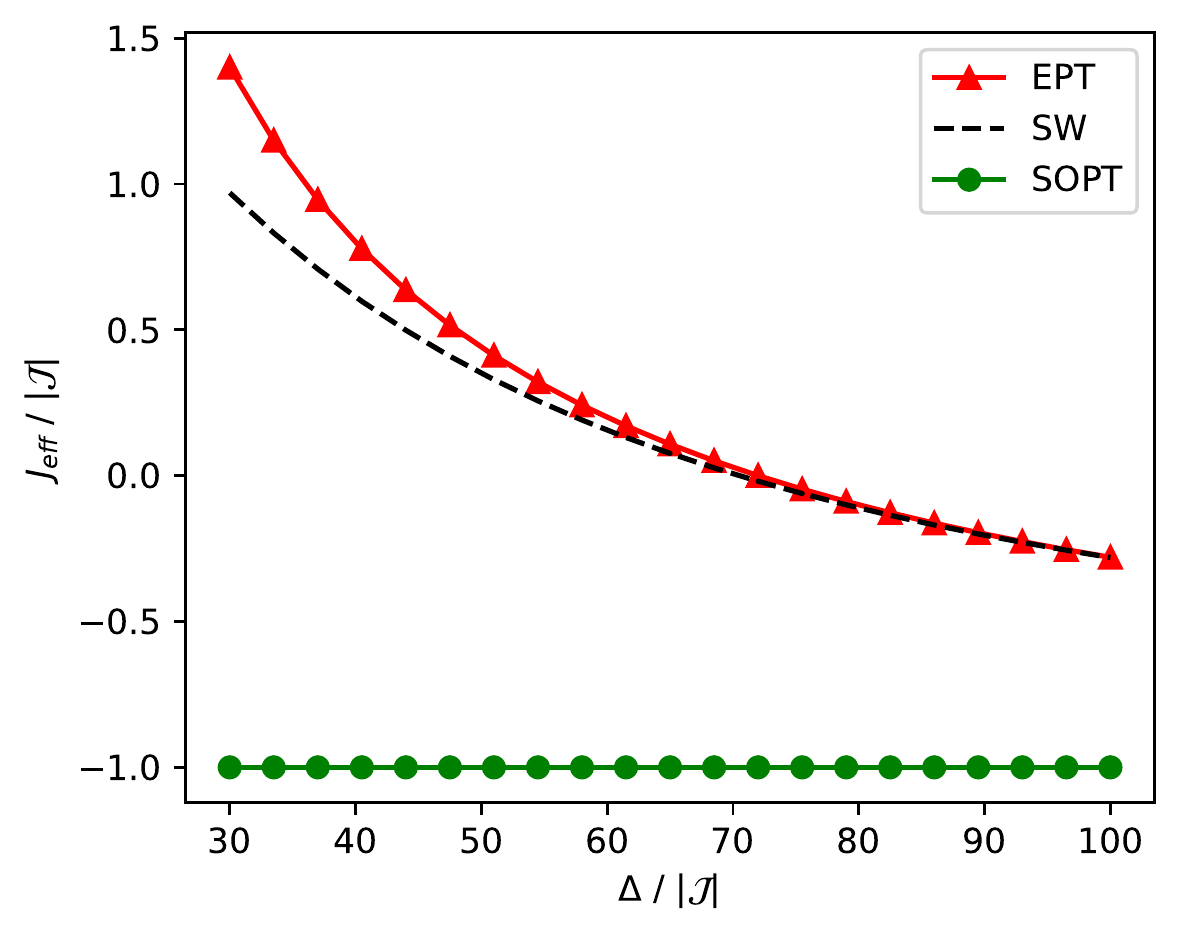}
    \caption{Comparison of the $J=1/2$ spin models extracted from a split $J=3/2$ model with various splittings. The plot shown the $\Delta$ dependence of the identical diagonal components of the matrix $J_{ab}$ in the Hamiltonian $H_{\text{eff}} = \sum_{\langle i,j\rangle} J_{ab} s^a_i s^b_j$. All off diagonal components vanish. The $J=3/2$ model was chosen such that $J_{ab} = \text{diag}(-1,-1,-1)$ under projection, with additional coupling $K^{(i)} = \text{diag}(0,6,6,6)$. All other couplings are taken to be 0. Notice that EPT at $\Delta = 36|\mathcal{J}|$ gives an antiferromagnetic Heisenberg interaction with equal magnitude to the ferromagnetic Heisenberg interaction found via projection.}
    \label{fig:H2H}
\end{figure*}

Figure \ref{fig:H2H} details the $\Delta$ dependence of this toy model for a particular choice of parameters. In particular, the parameters are chosen so that simple projection gives a ferromagnetic Heisenberg model, but extended perturbation theory gives an antiferromagnetic Heisenberg model with equal magnitude at $\Delta=36\mathcal{J}$. The figure contrasts the standard projection with the extended perturbation theory. It also shows the results of integrating out the exciton with a Schreiffer-Wolff transformation. The full Hamiltonian used for this Schreiffer-Wolff trasnformation is the $J=3/2$ toy model described above; therefore, unlike Figures \ref{fig:d1Fig} and \ref{fig:d2Fig}, it gives no information regarding the agreement between EPT and an underlying microscopic model. Instead, the similarity between the EPT and SW results indicate that second order perturbation theory is sufficient to reliably extract to physical effects of the exciton from the $J=3/2$ model. 

\subsection{Heisenberg to Kitaev}
Consider a naive Heisenberg ferromagnet or antiferromagnet with second order spin model given by Equation \ref{eqn:Heisenberg}. Such a material could have a significant Kitaev interaction in the presence of some $K^{(r)}$ couplings. For example, consider the couplings
\begin{equation}
    K^{(i)} = \begin{pmatrix}
    0 & 0 & 0 & 0 \\
    0 & \sqrt{\frac{\mathcal{K} \Delta}{2}} & 0 & 0 \\
    0 & 0 & \sqrt{\frac{\mathcal{K} \Delta}{2}} & 0 \\
    0 & 0 & 0 & -\mathcal{J} \sqrt{\frac{\Delta}{2\mathcal{K}}}
    \end{pmatrix}
\end{equation}
with $\mathcal{K}>0$. The extended perturbation theory including these couplings gives
\begin{equation}
    \delta J = \begin{pmatrix}
    -\mathcal{J} & 0 & 0 \\
    0 & -\mathcal{J} & 0 \\
    0 & 0 & \mathcal{K} 
    \end{pmatrix}.
\end{equation}
Thus, the true physical theory is 
\begin{equation}
\label{eqn:H2Kresult}
    J = \begin{pmatrix}
    0 & 0 & 0 \\
    0 & 0 & 0 \\
    0 & 0 & \mathcal{J} + \mathcal{K}
    \end{pmatrix}.
\end{equation}
The inter-doublet couplings generate Kitaev interactions.

\begin{figure*}
    \centering
    \includegraphics[width=0.8\textwidth]{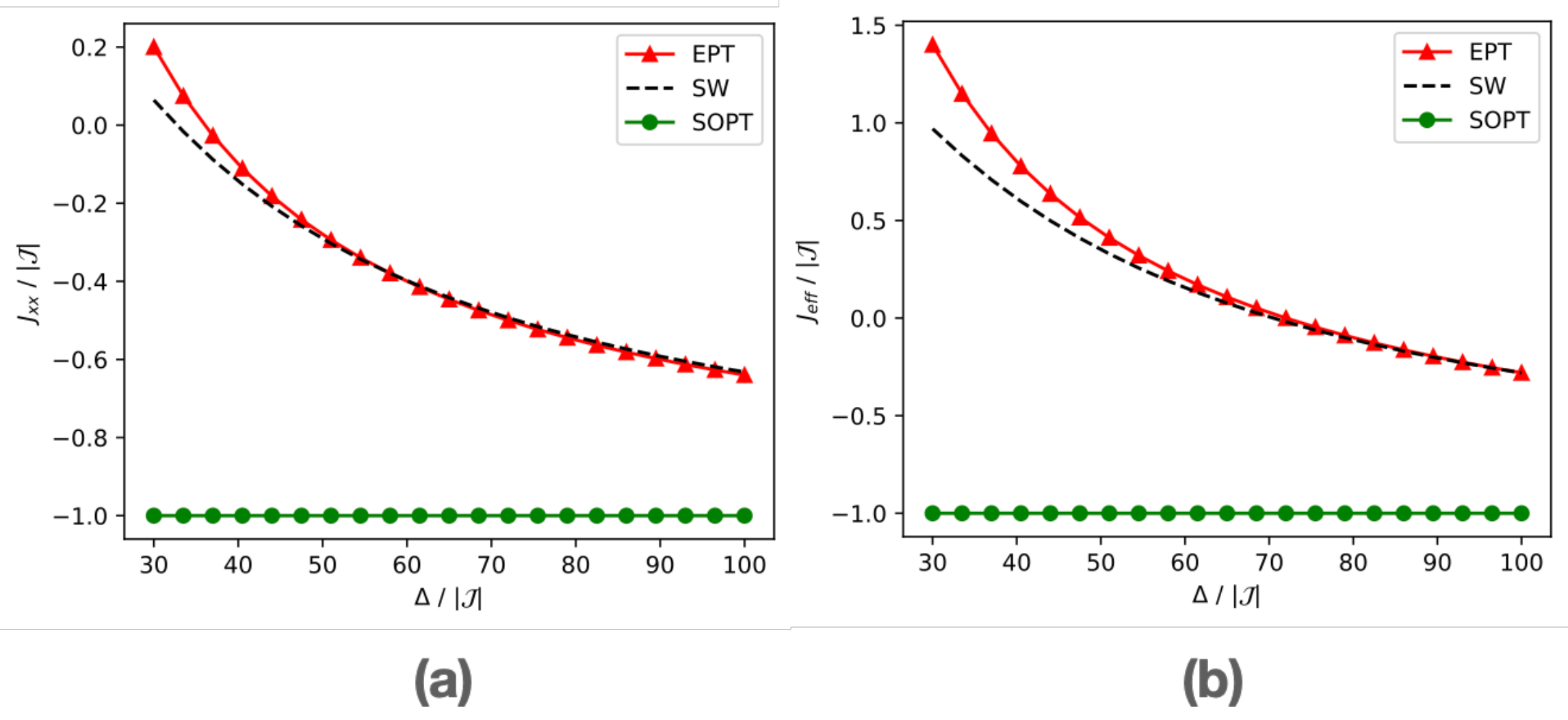}
    \caption{Comparison of the $J=1/2$ spin models extracted from a split $J=3/2$ model with various splittings. The subplots show the different elements of the matrix $J_{ab}$ in the Hamiltonian $H_{\text{eff}} = \sum_{\langle i,j\rangle} J_{ab} s^a_i s^b_j$. (a) $\Delta$ dependence of the $xx$-component, which is identical to the $yy$-component. (b) $\Delta$ dependence of the $zz$-component. All other components vanish. The $J=3/2$ model was chosen such that $J_{ab} = \text{diag}(-1,-1,-1)$ under projection, with additional coupling $K^{(i)} = \text{diag}(0,6,6,3)$. All other couplings are taken to be 0. Notice that EPT gives a pure Kitaev interaction at $\Delta = 36|\mathcal{J}|$.}
    \label{fig:H2K}
\end{figure*}

The various approaches to extracting the effective $J=1/2$ spin model from this $J=3/2$ model are contrasted in Figure \ref{fig:H2K}, in a similar manner to Figure \ref{fig:H2H}. The parameters are chosen such that the EPT methods results in Equation \ref{eqn:H2Kresult} at $\Delta=36|\mathcal{J}|$ with $\mathcal{J}<0$ and $\mathcal{K}=2|\mathcal{J}|$. Here again we see the agreement between the EPT and SW approaches implying that the second order perturbation theory is sufficient to reliably integrate out the exciton.

\subsection{Heisenberg to $K\Gamma$}
Extended perturbation theory can also lead to the development of off diagonal terms in the resulting spin theory. To demonstrate this consider another naive Heisenberg model which will turn into a $K\Gamma$ model with the inclusion of some inter-doublet couplings.

To produce a the $\Gamma$ interaction, we take the following couplings
\begin{equation}
    M^{(rr)} = \begin{pmatrix}
    0 & 0 & 0 & 0 \\
    0 & 0 & 0 & 0 \\
    0 & 0 & 0 & 2\sqrt{\left|\Gamma\right|\Delta}\\
    0 & 0 & 2\sqrt{\left|\Gamma\right|\Delta} & 0
    \end{pmatrix}
\end{equation}
and 
\begin{equation}
    M^{(ii)} = \begin{pmatrix}
    0 & 0 & 0 & 0 \\
    0 & 0 & 0 & -2\,\text{sign}\left(\Gamma\right)\sqrt{\left|\Gamma\right|\Delta} \\
    0 & 0 & 0 & \\
    0 & -2\,\text{sign}\left(\Gamma\right)\sqrt{\left|\Gamma\right|\Delta} & 0 & 0
    \end{pmatrix}.
\end{equation}
With these couplings, we generate the following correction to the effective spin model
\begin{equation}
    \delta J = \begin{pmatrix}
        -\left|\Gamma\right| & \Gamma & 0 \\
        \Gamma & -\left|\Gamma\right| & 0 \\
        0 & 0 & 0
    \end{pmatrix}.
\end{equation}
This generates the $\Gamma$ interaction. One can then use the $K^{(i)}$ to make the appropriate changes to the diagonal elements as in the previous subsection. Explicitly, the following coupling matrix does the trick
\begin{equation}
K^{(i)}  = \begin{pmatrix}
    0 & 0 & 0 & 0 \\
    0 & \sqrt{\frac{\mathcal{K} \Delta}{2}} & 0 & 0 \\
    0 & 0 & \sqrt{\frac{\mathcal{K} \Delta}{2}} & 0 \\
    0 & 0 & 0 & -\left(\mathcal{J}-\left|\Gamma\right|\right) \sqrt{\frac{\Delta}{2\mathcal{K}}}
\end{pmatrix}.
\end{equation}
With these couplings, the physical spin model as given by the extended perturbation theory, is 
\begin{equation}
\label{eqn:H2KGresult}
    J = \begin{pmatrix}
    0 & \Gamma & 0 \\
    \Gamma & 0 & 0 \\
    0 & 0 & \mathcal{J} + \mathcal{K}
    \end{pmatrix}.
\end{equation}

\begin{figure*}
    \centering
    \includegraphics[width=\textwidth]{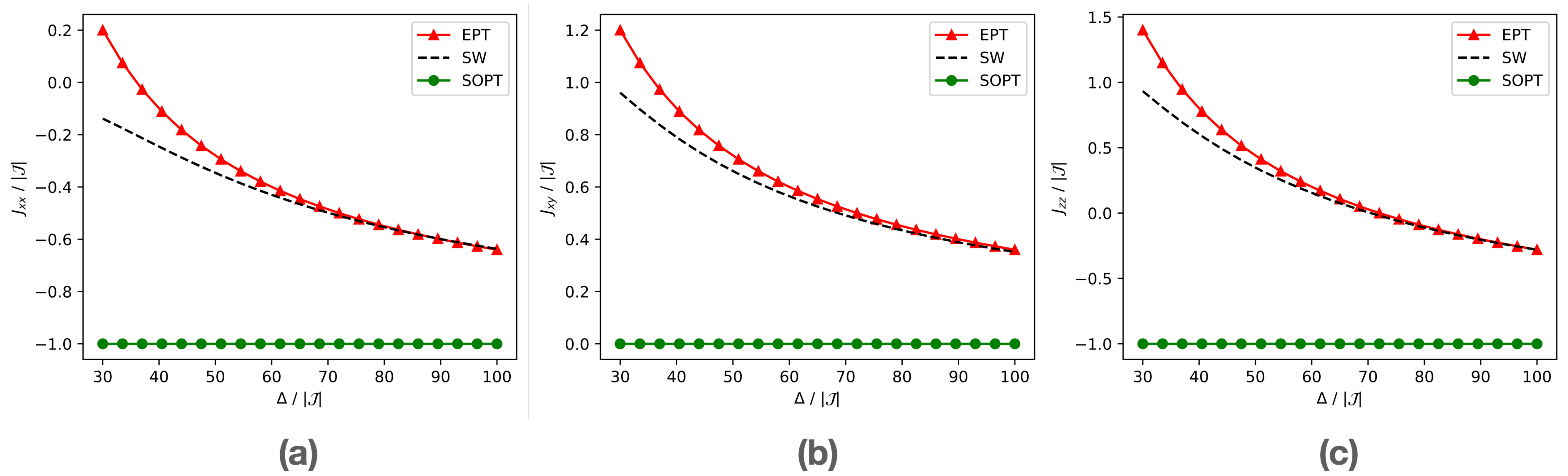}
    \caption{Comparison of the $J=1/2$ spin models extracted from a split $J=3/2$ model with various splittings. The subplots represent different elements of the matrix $J_{ab}$ in the Hamiltonian $H_{\text{eff}} = \sum_{\langle i,j\rangle} J_{ab} s^a_i s^b_j$. (a) $\Delta$ dependence of the $xx$-element, which is identical to the $yy$ element. (b) $\Delta$ dependence of the $zz$-element. (c) $\Delta$ dependence of the $xy$-element, which is identical to the $yx$-element. The remain elements have significantly smaller changes and may be seen in Figure \ref{fig:H2KG_2}. The $J=3/2$ model was chosen such that $J_{ab} = \text{diag}(-1,-1,-1)$ under projection, with additional coupling $K^{(i)} = \text{diag}(0,6,6,6)$, $M^{(rr)}_{23} = M^{(rr)}_{32} = 12$, and $M^{(ii)}_{13} = M^{(ii)}_{31} = -12$. All other couplings are taken to be 0. Notice that EPT gives a pure $K\Gamma$ interaction at $\Delta = 36|\mathcal{J}|$.}
    \label{fig:H2KG}
\end{figure*}

\begin{figure*}
    \centering
    \includegraphics[width=0.8\textwidth]{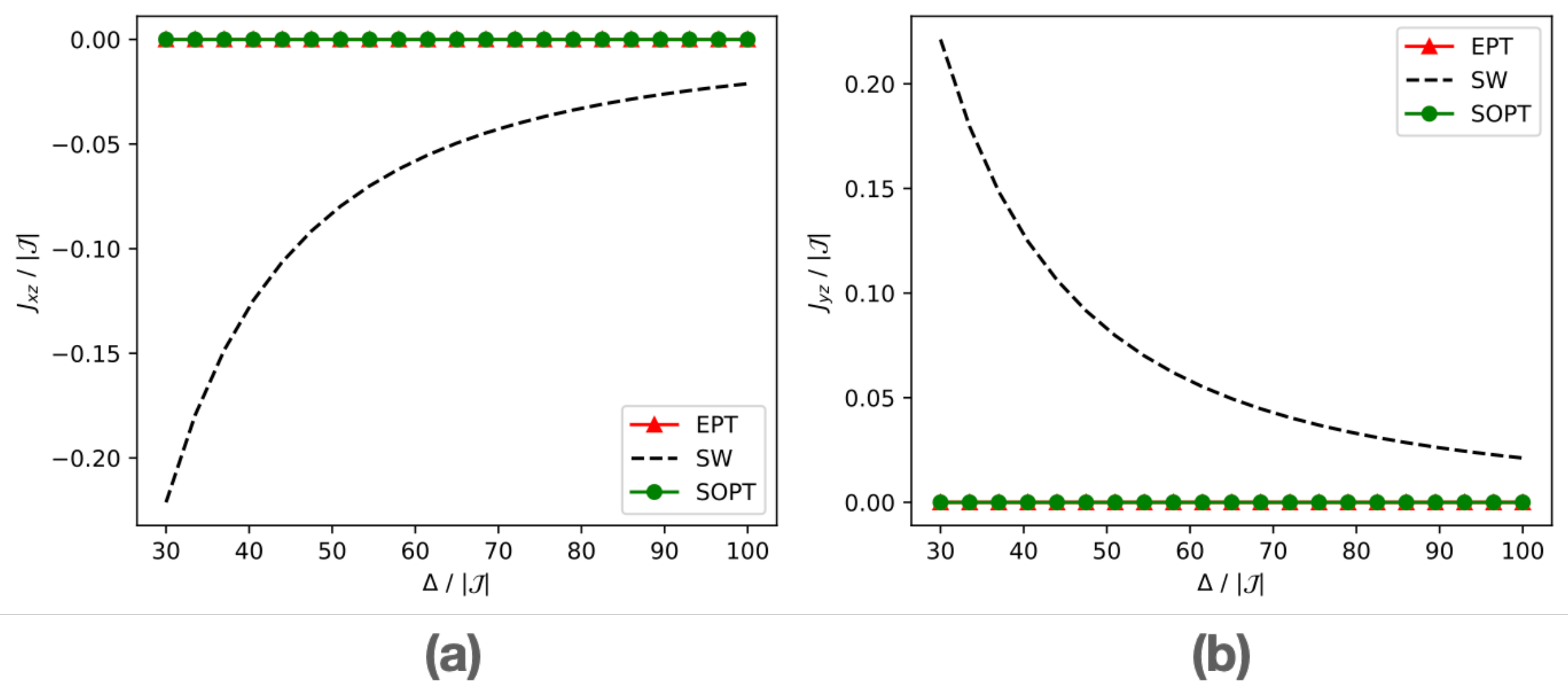}
    \caption{(a) Comparison of the $xz$-component of the model in Figure \ref{fig:H2KG}, which is identical to the $zx$-component. (b) Comparison of the $yz$-component of the model in Figure \ref{fig:H2KG}, which is identical to the $zy$-component.}
    \label{fig:H2KG_2}
\end{figure*}

In the spirit of Figures \ref{fig:H2H} and \ref{fig:H2K}, we plot the results of various methods of integrating out the exciton in Figure \ref{fig:H2KG}. The parameters were chosen such that EPT results in Equation \ref{eqn:H2KGresult} at $\Delta=36|\mathcal{J}|$ with $\mathcal{J}<0$, $\mathcal{K} = 2|\mathcal{J}|$, and $\Gamma=|\mathcal{J}|$. The similarity between the EPT and SW lines demonstrate once again that second order perturbation theory is sufficient to integrate out the excitons for the range of $\Delta$ we consider. However, in this case, as shown in Figure \ref{fig:H2KG_2}, the $J_{xz}$ and $J_{yz}$ components in the SW calculation do not vanish. While these effects are small compared to the other components, it indicates that at low $\Delta$ there is some other effect, likely involving mixing between the $M$ and $K$ terms. Since this mixing cannot occur at second order, this suggests that higher order terms will be necessary to consider at lower values of $\Delta$. We are only concerned with intermediate values of $\Delta$ where $J\ll \Delta$ still holds, so we are satisfied with the performance of second order perturbation theory and leave the study of higher order effects for future work.   

\section{Conclusion}
The extended perturbation theory described here is a simple and accurate technique for improving effective spin-1/2 models derived from second order perturbation theory in electron hoppings. By including the first excited multiplet on each site then integrating it out via a second perturbation step, the leading fourth order effects are included in the resulting Hamiltonian. 

The effectiveness of this approach was demonstrated clearly in Section \ref{} for the case of $d^1$ and $d^2$ systems where it accurately followed the results of non-perturbative Schreiffer-Wolff tranformations.

In addition, the results of our toy models in Section \ref{sec:ToyModels} demonstrate that this method can produce a wide variety of effects that are not included in the ordinary second order approach. 

In principle, one could take either perturbation step to higher order. If we call the approach presented here as a $2+2$ extended perturbation theory (owing to the fact that we take the second order results of each step), we could also consider a general $n+m$ extended perturbation theory. The examples above indicate that this is not necessary for either the realistic systems or the toy models we considered. In general, we suspect that $2+2$ is sufficient for most systems. More involved discussions of higher orders are left for future studies.

Based on the fact that this approach is simple to implement, is accurate in describing the physics, and can produce drastic results, it should be considered for future effective Hamiltonian searches in a wider setting, especially when the ordinary second order perturbation theory does not accurately describe the observed physics.

\section{Acknowledgements}
The authors wish to acknowledge the helpful contributions of Arijit Haldar in the early stages of this research. This work was supported by the Natural Sciences and Engineering Research Council of Canada.

\bibliography{d7Pert}

\appendix

\section{Extended Perturbation Theory Details}

In this appendix, we fill in the details in the derivation of Eqns. \eqref{KHamPT} and \eqref{MHamPT}. As mentioned in the main text, we assume that we have an interaction matrix $16 \times 16$ matrix $\mathcal{H}_{3/2}$ that contains all second order in $H_T$ contributions to the interactions between pseusodpins and excitons on the two sites. We split up the terms of $\mathcal{H}_{3/2}$ as follows:

\begin{equation}
\begin{split}
      \mathcal{H}_{3/2} = & H_0 + \mathcal{V}_J\\
      = & H_0 + \mathcal{V}_J^{(\eta\eta)} + \mathcal{V}_J^{(\eta\xi)} + \mathcal{V}_J^{(\xi\xi)}\\
      & + \mathcal{V}_J^{(\eta\tau)} + \mathcal{V}_J^{(\tau\tau)} + \mathcal{V}_J^{(\tau\xi)}
\end{split}
\end{equation}
Here $H_0$ is the matrix that splits the spectrum on each site, namely
\begin{equation}
H_0 = \Delta\left( \tau^0\otimes\eta^0 + \eta^0\otimes\tau^0+2~\tau^0\otimes\tau^0 \right).
\end{equation}
Each $\mathcal{V}_J^{(st)}$ denotes the collection of terms that couple $s$ and $t$ operators on each site. For completeness, we write each one out fully, but as stated in the main text only $\mathcal{V}_J^{(\eta\eta)}$, $\mathcal{V}_J^{(\eta\xi)}$, and $\mathcal{V}_J^{(\xi\xi)}$ can contribute to the effective $J=1/2$ Hamiltonian at second order. By our conventions, 
\begin{equation}
\begin{split}
\mathcal{V}_J^{(\eta\eta)} = J^{(\eta\eta)}_{ab} & \, \eta^a\otimes\eta^b,
\end{split}
\end{equation}
\begin{equation}
\begin{split}
\mathcal{V}_J^{(\eta\xi)} = K^{(r)}_{ab} & \left( \eta^a\otimes\xi_r^b + \xi_r^b \otimes\eta^a\right)\\ & + K^{(i)}_{ab} \left(\eta^a\otimes\xi_i^b + \xi_i^b\otimes\eta^a\right),
\end{split}
\end{equation}
\begin{equation}
\begin{split}
\mathcal{V}_J^{(\xi\xi)} = M^{(rr)}_{ab}&\, \xi_r^a\otimes \xi_r^b + M^{(ii)}_{ab}\, \xi_i^a\otimes\xi_i^b\\ & + M^{(ri)}_{ab} \left(\xi_r^a\otimes\xi_i^b + \xi_i^b\otimes\xi_r^a\right),
\end{split}
\end{equation}
\begin{equation}
\begin{split}
\mathcal{V}_J^{(\eta\tau)} =  J^{(\eta\tau)}_{ab} \left(\eta^a\otimes\tau^b + \tau^b\otimes\eta^a\right),
\end{split}
\end{equation}
\begin{equation}
\begin{split}
\mathcal{V}_J^{(\tau\tau)} = J^{(\tau\tau)}_{ab}\, \tau^a\otimes\tau^b ,
\end{split}
\end{equation}
and
\begin{equation}
\begin{split}
\mathcal{V}_J^{(\tau\xi)} = N^{(r)}_{ab}& \left( \tau^a\otimes\xi_r^b + \xi_r^b \otimes\tau^a\right)\\ & + N^{(i)}_{ab} \left(\tau^a\otimes\xi_i^b + \xi_i^b\otimes\tau^a\right).
\end{split}
\end{equation}
Here the $J$, $K$, $M$, and $N$ symbols represent $4\times4$ matrices of real-valued coefficients that determine the interaction strengths of all the possible interactions. The indices $a$ and $b$ are meant to be summed over $\{0,1,2,3\}$ according to Einstein summation notation, whereas $r$ and $i$ are merely labels for the $\xi_r$ and $\xi_i$ operators and should not be summed over. We assume inversion symmetry between the sites, making $J^{\eta\eta}$, $J^{\tau\tau}$, $M^{(rr)}$, and $M^{(ii)}$ symmetric matrices. We will assume for the sake of perturbation theory that $J,K,M,N \ll \Delta $.

With just $H_0$, the $\left|J_z\right|= 1/2$ subspace is entirely trivial; no interactions occur between sites and on each site the two states are perfectly degenerate. Therefore, our use of degenerate perturbation theory is justified. Let $P_0 = \eta^0 \otimes \eta^0$ denote the projector onto the subspace with $\left|J_z\right| = 1/2$ on both sites, and $P_1 = I-P_0$. At first order, we get 
\begin{equation}
\begin{split}
H_{\text{eff}}^{[1]} = & P_0 \mathcal{V}_J P_0\\
= & \mathcal{V}_J^{(\eta\eta)}.
\end{split}
\end{equation}
At this order, only $\mathcal{V}_J^{(\eta\eta)}$ can contributes. This reproduces the standard interaction Hamiltonian obtained through merely projecting to the lower doublet. 

To get our extended perturbation theory result, we look at the second order correction given by Eqn. \eqref{ptexchange}, 
\begin{equation}
H_{\text{eff}}^{[2]} = - P_0 \mathcal{V}_J P_1 \frac{1}{H_0} P_1 \mathcal{V}_J P_0
\end{equation}
With a little algebra one can show that this splits into two terms $H_{\text{eff}}^{[2]}  = H^{[2]}_K + H^{[2]}_M$ with
\begin{equation}
\label{eqn:dHprime_1}
H^{[2]}_K = - P_0 \mathcal{V}^{(\eta\xi)} P_1 \frac{1}{H_0} P_1 \mathcal{V}^{(\eta\xi)} P_0
\end{equation}
and
\begin{equation}
\label{eqn:dHprimeprime_1}
 H^{[2]}_M = - P_0 \mathcal{V}^{(\xi\xi)} P_1 \frac{1}{H_0} P_1 \mathcal{V}^{(\xi\xi)} P_0.
\end{equation} 

Notice that $P_0 \mathcal{V}^{(\eta\xi)} P_1 = P_0 \mathcal{V}^{(\eta\xi)}$ vanishes on all states except for those that have $\left|J_z\right| = 1/2$ on one site and  $\left|J_z\right| = 3/2$ on the other. Hence, the only non-vanishing contribution sees $1/H_0$ as $1/\Delta$. Using this we can simplify our equation greatly, 
\begin{equation}
\label{eqn:dHprime_2}
H^{[2]}_K = -\frac{1}{\Delta} P_0 \left(\mathcal{V}^{(\eta\xi)}\right)^2 P_0.
\end{equation}
Similarly, $P_0 \mathcal{V}^{(\xi\xi)} P_1 = P_0 \mathcal{V}^{(\eta\xi)}$ kills all states except for those with $\left|J_z\right| = 3/2$ on both sites, allowing the simplification
\begin{equation}
\label{eqn:dHprimeprime_2}
H^{[2]}_M = -\frac{1}{2\Delta} P_0 \left(\mathcal{V}^{(\xi\xi)}\right)^2 P_0.
\end{equation}

By writing the $\xi$ operators explicitly as tensor products of Pauli matrices and recalling that we defined lambda through $\sigma^a\sigma^b = \lambda^{abc}\sigma^c$, we can rewrite $H^{[2]}_K = \left(\delta J_K\right)_{ab} \sigma^a\otimes\sigma^b$ and $H^{[2]}_M = \left(\delta J_M\right)_{ab} \sigma^a\otimes\sigma^b$. Following this procedure gives
\begin{equation}
\label{eqn:dJK}
\begin{split}
(\delta J_K)_{ab} = -\frac{1}{2\Delta} & \left(K^{(r)}_{cd} - i K^{(i)}_{cd}\right)\left(K^{(r)}_{ef} + i K^{(i)}_{ef}\right)\\ & \times \left(\lambda_{cea}\lambda_{dfb} + \lambda_{ceb}\lambda_{dfa}\right)
\end{split}
\end{equation}
and
\begin{equation}
\label{eqn:dJM}
\begin{split}
(\delta J_M)_{ab} = -&\frac{1}{8\Delta}   \left(M^{(rr)}_{cd} - M^{(ii)}_{cd} - i M^{(ri)}_{cd} - i M^{(ri)}_{cd} \right)\\ &\times \left(M^{(rr)}_{ef} - M^{(ii)}_{ef} + i M^{(ri)}_{ef} + i M^{(ri)}_{ef}\right)\lambda_{cea}\lambda_{dfb}
\end{split}
\end{equation}
as presented in the main text. Therefore, we can describe our effective Hamiltonian as
\begin{equation}
\begin{split}
    H_{\text{eff}} = & \left(J_{\text{eff}}\right)_{ab} \sigma^a\otimes\sigma^b \\
    = & \left(J^{(\eta\eta)} + \delta J_K + \delta J_M \right)_{ab} \sigma^a\otimes\sigma^b.
\end{split}
\end{equation}

\section{Proof of surjectivity of the extended perturbation theory equations}
\label{app:A2A}
In this section, we prove the claim made in Section \ref{sec:ToyModels} that the extended perturbation theory can create any change in the spin model. In other words, Equations \eqref{KHamPT} and \eqref{MHamPT}, which determine $\delta J$, are surjective onto the set of symmetric $3\times 3$ matrices. We will prove this claim by starting with an arbitrary $\delta J$ and constructing $K^{(i)}$, $M^{(rr)}$, and $M^{(ii)}$ couplings that produce $\delta J$. An important feature of this construction is that it will only involve non-zero coefficients for couplings which preserve the time-reversal and exchange symmetries. Hence, there are no symmetry restrictions to finding these couplings in nature.

Consider an arbitrary symmetric $3\times 3$ matrix, $\delta J$. For concreteness, we label the elements of this matrix as
\begin{equation}
    \delta J = \begin{pmatrix}
    k1 & m3 & m2 \\
    m3 & k2 & m1 \\
    m2 & m1 & k3
    \end{pmatrix}.
\end{equation}
Here we have been intentionally suggestive with our labels. Indeed the off-diagonal $m_i$ elements will be set by our choice of $M^{(rr)}$ and $M^{(ii)}$, then the diagonal will be set by our choice of $K^{(i)}$. 

Starting with the off-diagonal elements, the simplest case is that $m_1=m_2=m_3=0$. In this case, we can set $M^{(rr)} = M^{(ii)} = 0$ and move on to dealing with the diagonal elements. Otherwise, suppose $m_i\neq 0$ for some $i\in \{1,2,3\}$. It will be useful to define the following matrix-valued functions
\begin{equation}
    \begin{split}
        M_1(a,b) = & \begin{pmatrix}
            0 & 0 & 0 & 0 \\
            0 & b & 0 & 0 \\
            0 & 0 & 0 & -a \\
            0 & 0 & -a & 0
        \end{pmatrix}\\
        M_2(a,b) = & \begin{pmatrix}
            0 & 0 & 0 & 0 \\
            0 & 0 & 0 & -a \\
            0 & 0 & b & 0 \\
            0 & -a & 0 & 0
        \end{pmatrix}\\
        M_3(a,b) = & \begin{pmatrix}
            0 & 0 & 0 & 0 \\
            0 & 0 & -a & 0 \\
            0 & -a & 0 & 0 \\
            0 & 0 & 0 & b \\
        \end{pmatrix},
    \end{split}
\end{equation}
where $a$ and $b$ are real numbers. In what follows, all arithmetic involving indices is modulo 3. 

Consider setting $M^{(rr)} = M_{i+1}(a,b)$, $M^{(ii)} = M_{i+2}(c,d)$, $M^{(ri)} = 0$, and $M^{(ir)}=0$, for some real numbers $a$, $b$, $c$, and $d$. Plugging these into Equation \ref{eqn:dJM} and equating the off-diagonal elements to the off-diagonal components of $\delta J$, one finds the following system of equations
\begin{equation}
    \begin{split}
        m_i = & -\frac{ac}{4\Delta}\\
        m_{i+1} = & \frac{ab}{4\Delta}\\
        m_{i+2} = & \frac{cd}{4\Delta}.
    \end{split}
\end{equation}
A solution to this system of equations is given by 
\begin{equation}
    \begin{split}
        a = & -2\text{sign}(m_i)\sqrt{\left|m_i\right|\Delta}\\
        b = & -2\text{sign}(m_i) m_{i+1} \sqrt{\frac{\Delta}{\left|m_i\right|}}\\
        c = & 2\sqrt{\left|m_i\right|\Delta}\\
        d = & 2 m_{i+2} \sqrt{\frac{\Delta}{\left|m_i\right|}}.
    \end{split}
\end{equation}
Hence, to have $\delta J_M$ have the desired off-diagonal elements, define
\begin{equation}
\begin{split}
   & M^{(rr)} = \\
    & M_{i+1} \left(-2\text{sign}(m_i)\sqrt{\left|m_i\right|\Delta} , -2\text{sign}(m_i) m_{i+1} \sqrt{\frac{\Delta}{\left|m_i\right|}}\right)
\end{split}
\end{equation}
and
\begin{equation}
    M^{(ii)} = M_{i+2} \left(2\sqrt{\left|m_i\right|\Delta},2 m_{i+2} \sqrt{\frac{\Delta}{\left|m_i\right|}}\right).
\end{equation}

In general, the resulting $\delta J_M$ will have non-zero diagonal elements\footnote{The exact form of these elements depends on the choice of index $i$ when defining $M^{(rr)}$ and $M^{(ii)}$. The values of these elements will have little bearing on the procedure for finding $K^{(i)}$, so we do not give it explicitly.}, so to fully reproduce the desired $\delta J$, we need to find $K^{(r)}$ and $K^{(i)}$ such that 
\begin{equation}
    \delta J_K = \begin{pmatrix}
    k_1 - \delta J_{M,11} & 0 & 0 \\
    0 & k_2 - \delta J_{M,22} & 0 \\
    0 & 0 & k_3 - \delta J_{M,33}
    \end{pmatrix}.
\end{equation}
For simplicity of notation, we define $\tilde{k}_i = k_i - \delta J_{M,ii}$. Consider setting $K^{(r)}=0$ and 
\begin{equation}
    K^{(i)} = \begin{pmatrix}
    a_0 & 0 & 0 & 0 \\
    0 & a_1 & 0 & 0 \\
    0 & 0 & a_2 & 0 \\
    0 & 0 & 0 & a_3
    \end{pmatrix},
\end{equation}
with $a_0,a_1,a_2,a_3\in\mathbb{R}$. Evaluating $\delta J_K$ and equating with the desired form gives the following system of equations
\begin{equation}
\label{eqn:Ksystem}
    \begin{split}
        a_0a_1-a_2a_3 =& -\frac{2}{\Delta}\tilde{k}_1 \equiv \kappa_1\\
        a_0a_2-a_3a_1 =& -\frac{2}{\Delta}\tilde{k}_2 \equiv \kappa_2\\
        a_0a_3-a_1a_2 =& -\frac{2}{\Delta}\tilde{k}_3 \equiv \kappa_3.
    \end{split}
\end{equation}
A useful trick for finding a general solution to these equations is to set 
\begin{equation}
    \begin{split}
        a_0 = & b-c\\
        a_1 = & b+c\\
        a_2 = & f+e\\
        a_3 = & f-e
    \end{split}
\end{equation}
for some real numbers $b$, $c$, $e$, and $f$. This changes the Equations \ref{eqn:Ksystem} to
\begin{equation}
\label{eqn:Ksystem_alt}
    \begin{split}
        b^2 - c^2 - e^2 + f^2 = & \kappa_1 \\
        -2 ce + 2 bf = & \kappa_2\\
        -2 ce - 2 bf = & \kappa_3
    \end{split}
\end{equation}
The last two equations imply that
\begin{equation}
\label{eqn:cebf}
\begin{split}
    ce = & -\frac{1}{4}\left(\kappa_2 + \kappa_3\right) \equiv \kappa_+\\
    bf = & \frac{1}{4}\left(\kappa_2-\kappa_3\right) \equiv \kappa_-.
\end{split}
\end{equation}
We are free to choose our solution such that $e,f\neq 0$, in which case we may combine Equations \ref{eqn:cebf} with the first equation of \ref{eqn:Ksystem_alt} to find
\begin{equation}
    f^2 + \frac{\kappa_+^2}{f^2} - e^2 - \frac{\kappa_-^2}{e^2} = \kappa_1.
\end{equation}
This equation can be solved with
\begin{widetext}
\begin{equation}
    f = \begin{cases}
    \sqrt{\kappa_1 + \abs{\kappa_+} + \abs{\kappa_-} + \sqrt{\left(\kappa_1 + \abs{\kappa_+}+\abs{\kappa_-}\right)^2 - 4\kappa_-^2}}& \kappa_1 \geq 0\\
    \sqrt{\frac{\abs{\kappa_1}}{2} + \abs{\kappa_+} + \abs{\kappa_-} + \sqrt{\left(\frac{\abs{\kappa_1}}{2} + \abs{\kappa_+}+\abs{\kappa_-}\right)^2 - 4\kappa_-^2}}& \kappa_1 < 0
    \end{cases}
\end{equation}
and
\begin{equation}
    e = \begin{cases}
    \sqrt{\frac{\kappa_1}{2} + \abs{\kappa_+} + \abs{\kappa_-} + \sqrt{\left(\frac{\kappa_1}{2} + \abs{\kappa_+}+\abs{\kappa_-}\right)^2 - 4\kappa_+^2}}& \kappa_1 \geq 0\\
    \sqrt{\abs{\kappa_1} + \abs{\kappa_+} + \abs{\kappa_-} + \sqrt{\left(\abs{\kappa_1} + \abs{\kappa_+}+\abs{\kappa_-}\right)^2 - 4\kappa_+^2}}& \kappa_1 < 0
    \end{cases}.
\end{equation}
\end{widetext}
These solutions can then be used to find $b = \kappa_- / f$ and $c = \kappa_+ / e$, and subsequently $a_0 = b-c$, $a_1 = b+c$, $a_2 = e+f$, and $a_3 = e-f$. The resulting $K^{(i)}$ gives the desired $\delta J_K$, completing the proof.

\section{Relations between operators}
\label{app:bases}
Here we provide useful details about the operators used in the main text to describe $J=3/2$ degrees of freedom. The definition of the multipole basis in terms of the dipole operators, which form a $J=3/2$ representation of the $\mathfrak{su}(2)$ algebra, is presented in Table \ref{tab:1}. These definitions are well-known; we include them for the sake of completeness and transparency with our conventions. The relationship between this basis and the basis defined in Equations \eqref{eqn:basisdef1} and \eqref{eqn:basisdef2} is given in Table \ref{tab:2}. 

\renewcommand{\arraystretch}{3.0}
\setlength{\tabcolsep}{20pt}
\begin{table*}[t]
  \centering
  \begin{tabular}{|c|c|c|c|}
  \hline
    Moment & Symmetry & Symbol & Expression \\
    \hline
    \multirow{3}{*}{Dipole} & \multirow{3}{*}{$T_1$} & $J_x$ &  \\
                              &                          & $J_y$ &  \\
                              &                          & $J_z$ &  \\
    \hline 
    \multirow{5}{*}{Quadrupole} & \multirow{3}{*}{$T_2$} & $Q_{yz}$ & $\dfrac{1}{\sqrt{3}} \overline{J_y J_z}$ \\
                                  &                          & $Q_{zx}$ & $\dfrac{1}{\sqrt{3}} \overline{J_z J_x}$ \\
                                  &                          & $Q_{xy}$ & $\dfrac{1}{\sqrt{3}} \overline{J_x J_y}$ \\ \cline{2-4}
                                  & \multirow{2}{*}{$E$} & $Q_{x^2-y^2}$ & $\dfrac{1}{\sqrt{3}} \left( J_x^2 - J_y^2 \right)$ \\
                                  &                        & $Q_{z^2}$ & $\dfrac{1}{3} \left( 3J_z^2 - \mathbf{J}^2 \right)$ \\ 
    \hline
    \multirow{7}{*}{Octupole}   & $A_2$                  & $T_{xyz}$ & $\dfrac{2}{3\sqrt{3}} \overline{J_xJ_yJ_z}$ \\ \cline{2-4}
                                  & \multirow{3}{*}{$T_1$} & $T_x^a$ & $\dfrac{2}{3} \left( J_x \right)^3 - \dfrac{1}{3}\left( \overline{J_x \left( J_y \right)^2} + \overline{ \left( J_z \right)^2 J_x} \right)$ \\
                                  &                          & $T_y^a$ & $\dfrac{2}{3} \left( J_y \right)^3 - \dfrac{1}{3}\left( \overline{J_y \left( J_z \right)^2} + \overline{ \left( J_x \right)^2 J_y} \right)$ \\
                                  &                          & $T_z^a$ & $\dfrac{2}{3} \left( J_z \right)^3 - \dfrac{1}{3}\left( \overline{J_z \left( J_x \right)^2} + \overline{ \left( J_y \right)^2 J_z} \right)$ \\ \cline{2-4}
                                  & \multirow{3}{*}{$T_2$} & $T_x^b$ & $\dfrac{2}{3 \sqrt{3}} \left( \overline{J_x \left( J_y \right)^2} - \overline{ \left( J_z \right)^2 J_x} \right)$ \\
                                  &                          & $T_y^b$ & $\dfrac{2}{3 \sqrt{3}} \left( \overline{J_y \left( J_z \right)^2} - \overline{ \left( J_x \right)^2 J_y} \right)$ \\
                                  &                          & $T_z^b$ & $\dfrac{2}{3 \sqrt{3}} \left( \overline{J_z \left( J_x \right)^2} - \overline{ \left( J_y \right)^2 J_z} \right)$ \\ 
    \hline
  \end{tabular}
  \caption{The definition of the various multipole operators for a $J=3/2$ system in terms of the dipole operators. These operators form a useful basis of $\mathfrak{su}(4)$.}
  \label{tab:1}
\end{table*}

\begin{table*}[t]
  \centering
  \begin{tabular}{|c|c|c|}
  \hline
  Type & Symbol & Expression \\
  \hline
  \multirow{4}{*}{Lower Doublet} & $\eta^0$ & $-\dfrac{1}{2} Q_{z^2} + \dfrac{1}{2}$ \\
  & $\eta^x$ & $\dfrac{2}{5} J_x + \dfrac{3}{10} T_x^a + \dfrac{\sqrt{3}}{4} T_x^b$ \\
  & $\eta^y$ & $\dfrac{2}{5} J_y + \dfrac{3}{10} T_y^a - \dfrac{\sqrt{3}}{4} T_y^b$ \\
  & $\eta^z$ & $\dfrac{1}{5} J_z + \dfrac{3}{5} T_z^a$ \\
  \hline
  \multirow{4}{*}{Upper Doublet} & $\tau^0$ & $\dfrac{1}{2} Q_{z^2} + \dfrac{1}{2}$ \\
  & $\tau^x$ & $\dfrac{1}{2} T_x^a - \dfrac{\sqrt{3}}{4} T_x^b$ \\
  & $\tau^y$ & $-\dfrac{1}{2} T_y^a - \dfrac{\sqrt{3}}{4} T_y^b$ \\
  & $\tau^z$ & $-\dfrac{3}{5} J_z + \dfrac{1}{5} T_z^a$ \\
  \hline 
  \multirow{8}{*}{Mixing Between Doublets} & $\xi_r^0$ & $\dfrac{\sqrt{6}}{5} J_x - \dfrac{\sqrt{3}}{5\sqrt{2}} T_x^a - \dfrac{1}{2\sqrt{2}} T_x^b$ \\
  & $\xi_r^x$ & $\dfrac{1}{\sqrt{2}} Q_{x^2-y^2}$ \\
  & $\xi_r^y$ & $\dfrac{1}{\sqrt{2}} Q_{xy}$ \\
  & $\xi_r^z$ & $\dfrac{1}{\sqrt{2}} Q_{zx}$ \\ \cline{2-3}
  & $\xi_i^0$ & $-\dfrac{1}{\sqrt{2}} Q_{yz}$ \\
  & $\xi_i^x$ & $-\dfrac{1}{\sqrt{2}} T_{xyz}$ \\
  & $\xi_i^y$ & $\dfrac{1}{\sqrt{2}} T_z^b$ \\
  & $\xi_i^z$ & $-\dfrac{\sqrt{6}}{5} J_y + \dfrac{\sqrt{3}}{5\sqrt{2}} T_y^a - \dfrac{1}{2\sqrt{2}} T_y^b$ \\
  \hline
  \end{tabular}
  \caption{The relation between the magnetic multipole operator basis for a $J=3/2$ system and the basis introduced in Equations \eqref{eqn:basisdef1} and \eqref{eqn:basisdef2}. Here we have included the identity operator with the multipole basis, so these basis describe the 16-dimensional space, $\mathfrak{u}(4)$.}
  \label{tab:2}
\end{table*}

\end{document}